\documentclass[prd,aps,nofootinbib,floatfix,10pt]{revtex4}
\usepackage{amsmath,graphicx,epsfig,amssymb,dsfont,mathtools,}
\usepackage[usenames]{color}
\usepackage{ulem} 
\usepackage{cancel}
\usepackage{bigstrut}
\usepackage{slashed}
\usepackage{multirow}
\usepackage{makecell}
\usepackage{verbatim}
\usepackage{amsmath,graphicx,color,epsfig}
\usepackage{longtable}
\usepackage{subfigure}
  \allowdisplaybreaks[1]

\allowdisplaybreaks

\begin{document}
\title{S-wave contributions in $\bar B_s^0\to (D^0,\bar D^0)\pi^+\pi^- $ within perturbative QCD approach}
\author{Ye Xing$^1$~\footnote{Email:xingye\_guang@sjtu.edu.cn}, Zhi-Peng Xing$^1$~\footnote{Email:zpxing@sjtu.edu.cn}
}

\affiliation{
 $^1$ INPAC,  SKLPPC, MOE Key Laboratory for Particle Physics, School of Physics and Astronomy, \\ Shanghai Jiao Tong University, Shanghai  200240, China}

\begin{abstract}
The $\bar B_s^0\to (D^0,\bar D^0) \pi^+\pi^-$ is induced by the $b\to c \bar us$/$b \to u\bar cs$ transition,  and can interfere if  a CP-eigenstate $D_{\rm CP}$ is formed.  The interference contribution is sensitive to the CKM angle $\gamma$. In this work, we study S-wave $\pi^+\pi^-$ contributions to the process in the perturbative QCD factorization. Under the factorization framework, we adopt  two-meson light-cone distribution amplitudes, whose normalization is parametrized by  the S-wave time-like two-pion form factor with the resonance contributions from $f_0(500)$, $f_0(980),f_0(1500),f_0(1790)$.  We find the branching ratios of $\bar B_s^0\to (D^0,\bar D^0) (\pi^+\pi^-)_S$ can reach the order of $10^{-6}$, and significant interferences exist in $\bar B_s^0\to D_{CP} (\pi^+\pi^-)_S$.   The future measurement can not only provide useful constraints on the CKM angle $\gamma$ but is also helpful to explore the multi-body decay mechanism of heavy mesons.
\end{abstract}

\maketitle

\section{Introduction}
\label{sec:introduction}

In recent years,
three-body hadronic $B$/$B_s$ meson decays have attracted great attentions on the experimental side~\cite{Aaij:2013pua,Aaij:2018rol,Aaij:2019ipm}. 
These processes are capable to  provide new sources to study the phenomenology in the Standard Model and probe the new physics effects.
For instance, LHCb Collaboration has measured sizable direct CP asymmetries in the various phase space of three-body $B$ decays~\cite{Aaij:2013bla,Aaij:2013sfa}. In addition, they  are also valuable  for us to understand the mechanism for multi-body heavy meson decays.

On the theoretical side,
the perturbative QCD(PQCD) approach, based on the $k_T$ factorization,  has been applied to analyze the $B$/$B_s$ semi-leptonic and  two-body decays processes~\cite{Yeh:1997rq,Li:2003yj,Li:1994iu,Ali:2007ff,Kurimoto:2002sb,Li:2012cfa,Li:2008tk,Kim:2013ria,Wang:2012ab,Cheng:2005nb,Li:2012nk,Lu:2018cfc,Li:2009wq,Wang:2015vgv,Li:2010nn,Li:2003az,Li:2004ep,Wang:2012ie,Lu:2011jm,Colangelo:2010bg,Wang:2006ria,Lu:2002ny,Lu:2000em,Lu:2018obb}. The PQCD approach has also been used  to study   three-body decays~\cite{Chen:2002th,Chen:2004az,Cheng:2013dua,Li:2014fla,Li:2017obb,Cheng:2014uga,Shi:2017pgh,Shi:2015kha,Wang:2015paa,Meissner:2013pba}.   Generally,  the multi-scale decay amplitude might  be written as  as a convolution, including the nonperturbative wave functions, hard kernel at the intermediate scale and   short-distance Wilson coefficients. The factorization  is greatly simplified if two of the final hadrons move collinearly. In this case, the three-body decays are reduced to quasi-two-body processes. Therefore, nonperturbative wave functions include two-meson light-cone distributions, which contain both resonant and nonresoant contributions. For instance,
the measurement of LHCb~\cite{Aaij:2013sfa} of $B_s\to J/\psi (\pi^+\pi^-)_S$ supports that the resonances $f_0(500)$,  $f_0(980),f_0(1500),f_0(1790)$ of the S-wave $\pi\pi$-pair are dominant, which is confirmed by the theoretical calculation in the frame of PQCD~\cite{Wang:2015uea,Ma:2016csn,Ma:2017idu,Li:2015tja,Meissner:2013hya}.  In this work, we will focus on the $\bar B_s^0\to D^0(\bar D^0) \pi^+\pi^-$, and include the  $B_s \to D (f_0(500)+f_0(980)+f_0(1500)+f_0(1790) )\to D[(\pi^+\pi^-)_S]$ contributions. More explicitly,  a Breit-Wigner(BW) model will be used for the resonance $f_0(500)$, $f_0(1500), f_0(1790)$~\cite{Aaij:2014emv} and Flatt\'{e} Model is adopted for  the resonance $f_0(980)$~\cite{Flatte:1976xv}.
The $\bar B_s^0\to D^0(\bar D^0) \pi^+\pi^-$, with CP eigenstate containing  the interference of $b\to c\bar u s$ ($b\to u\bar c s$) amplitude, is sensitive to the angle $\gamma$ of the CKM Unitarity Triangle whose precise measurement is one of the primary objectives in  flavour physics.


This paper is organized as follow: In Sec.II, we introduce the wave functions of $B_s$, $D$ and two pion mesons in turn, while Sec.III contains our perturbative calculation within the PQCD framework. In Sec.IV, we study the numerical results, and a conclusion is presented in the last section.

\section{Wave functions}
\label{sec:wave functions}

In general, wave function $\Phi_{\alpha\beta}$ with Dirac indices $\alpha,\beta$ can be decomposed into 16 independent components, $I_{\alpha\beta},\gamma^{\mu}_{\alpha\beta},(\gamma^{\mu}\gamma^5)_{\alpha\beta},\gamma^5_{\alpha\beta},\sigma^{\mu\nu}_{\alpha\beta}$.
For the pseudoscalar $B_s$ meson, the light-cone matrix element is defined as
\begin{eqnarray}
&&\int_{0}^1 \frac{d^4 z}{(2\pi)^4} e^{i k_1\cdot z} \langle0|b_{\alpha}(0)\bar q_{\beta}(z)|\bar B_s(P_{B_s})\rangle
=\frac{i}{\sqrt{2N_c}}\Bigg\{ ({P\!\!\!\!/}_{B_s}+m_{B_s})\gamma_5\Big[\phi_{B_{s}}(k_1)+\frac{n\!\!\!/-v\!\!\!/}{\sqrt{2}}\bar{\phi}_{B_{s}}(k_1)\Big]\Bigg\}_{\alpha\beta},
\end{eqnarray}
where the light-cone vectors $\textbf{n}=(1,0,0_T)$ and $\textbf{v}=(0,1,0_T)$. The two independent structures in $B_s$ meson light cone distribution amplitudes obey the following normalization conditions.
\begin{eqnarray}
\int \frac{d^4k_1}{(2\pi)^4}\phi_{B_s}(k_1)=\frac{f_{B_s}}{2\sqrt{2N_c}}, \int \frac{d^4k_1}{(2\pi)^4}\bar{\phi}_{B_s}(k_1)=0,
\end{eqnarray}
with $f_{B_s}$ as the decay constant of $B_s$ meson. Since the contribution of $\bar{\phi}_{B_s}(k_1)$ is numerically small~\cite{Lu:2002ny}, we neglect it and keep only $\phi_{B_s}(k_1)$ in the above equation. In   momentum space the light cone matrix of $B_s$ meson can be expressed as follows:
\begin{eqnarray}
\Phi_{B_s}=\frac{i}{\sqrt{2N_c}}({P\!\!\!\!/}_{B_s}+m_{B_s})\gamma_5 \phi_{B_s}(k_1).
\end{eqnarray}
Usually the hard part is independent of $k^+$ or/and $k^-$, thus one can  integrate one of them out from $\phi_{B_s}(k^+,k^-,k_{\perp})$. With $b$ as the conjugate space coordinate of $k_{\perp}$, we can express $\phi_{B_s}(x,k_{\perp})$ in $b$-space by
\begin{eqnarray}
\Phi_{B_s}(x,b)_{\alpha\beta}=\frac{i}{\sqrt{2N_c}}\Big[({P\!\!\!\!/}_{B_s}+m_{B_s})\gamma_5\Big]_{\alpha\beta} \phi_{B_s}(x,b),
\end{eqnarray}
where $x$ is the momentum fraction of the light quark in $B_s$ meson. In this paper, we adopt the following expression for $\phi_{B_s}(x,b)$
\begin{eqnarray}
\phi_{B_s}(x,b)=N_{B_s} x^2 (1-x)^2 {\rm exp}\Big[-\frac{m_{B_s}^2 x^2}{2\omega_b^2}-\frac{(\omega_b b)^2}{2}\Big],
\end{eqnarray}
with $N_{B_s}$ the normalization factor, which is determined by equation at $b=0$. In our calculation, we adopt $\omega_b=(0.50\pm0.05){\rm GeV}$ and $f_{B_s}=(0.23\pm0.03){\rm GeV}$~\cite{Ali:2007ff}, from which we determine the $N_{B_s}=63.58$.

The wave function of the charmed D meson, treated as the heavy-light system, is defined by the light cone matrix element as follows~\cite{Kurimoto:2002sb}:
\begin{eqnarray}
&&\int_{0}^1 \frac{d^4 z}{(2\pi)^4} e^{i k_2\cdot z} \langle 0|\bar c_{\alpha}(0) q_{\beta}(z)|\bar D^0(P_D))\rangle
=-\frac{i}{\sqrt{2N_c}}\Bigg\{ ({P\!\!\!\!/}_{D}+m_{D}^0)\gamma_5 \phi_{D}(k_2)\Bigg\}_{\beta\alpha},
\end{eqnarray}
which satisfies the normalization
\begin{eqnarray}
\int \frac{d^4k_2}{(2\pi)^4}\phi_{D}(k_2)=\frac{f_{D}}{2\sqrt{2N_c}}.  
\end{eqnarray}
Here $f_D$ is  the decay constant,  and  the chiral D meson mass is taken as $m^0_D=\frac{m_D^2}{m_c+m_d}=m_D+\mathcal{O}(\Lambda)$. For the numerical calculation, we adopt the  parametrization~\cite{Keum:2003js},
\begin{eqnarray}
\phi_{D}(x_2,b_2)=\frac{f_{D}}{2\sqrt{2N_c}}6x_2(1-x_2)[1+C_D (1-2x_2)]{\rm exp}\Big[-\frac{\omega_D^2 b_2^2}{2}\Big], 
\end{eqnarray}
with the free shape parameter $C_D$ taken as $C_D=0.5\pm0.1$, $f_D$, $\omega_D$ read as $f_D=0.221\pm 0.018$ and $\omega_D=0.1$, respectively~\cite{Kim:2013ria}.

Then  the S-wave two-pion distribution amplitudes is given as~\cite{Meissner:2013hya}
\begin{eqnarray}
\Phi_{\pi\pi}^{S-wave}=\frac{1}{\sqrt{2N_c}}[p\!\!\!/ \Phi_{\pi\pi}(z,\xi,m_{\pi\pi}^2)+m_{\pi\pi} \Phi_{\pi\pi}^s(z,\xi,m_{\pi\pi}^2)+m_{\pi\pi}(n\!\!\!/v\!\!\!/-1)\Phi_{\pi\pi}^T(z,\xi,m_{\pi\pi}^2)],
\end{eqnarray}
where $z$ is the momentum fraction carried by the spectator positive quark,  $\Phi_{\pi\pi}$, $\Phi_{\pi\pi}^s$ and $\Phi_{\pi\pi}^T$ are twist-2 and twist-3 distribution amplitudes. $m_{\pi\pi}$ is the invariant mass of the pion pair. We consider the two-pion system move in the \textbf{n} direction. $\xi$ as the momentum fraction of $\pi^+$ in pion pair.
The asymptotic forms are parameterized as~\cite{Mueller:1998fv,Diehl:1998dk,Polyakov:1998ze}
\begin{eqnarray}
\Phi_{\pi\pi}=\frac{F_s(m_{\pi\pi}^2)}{2\sqrt{2N_c}} a_2 6 z(1-z) 3(2z-1), \ \Phi_{\pi\pi}^s=\frac{F_s(m_{\pi\pi}^2)}{2\sqrt{2N_c}},\
\Phi_{\pi\pi}^T=\frac{F_s(m_{\pi\pi}^2)}{2\sqrt{2N_c}}(1-2z).
\end{eqnarray}
Here, $F_s(m_{\pi\pi}^2)$ and $a_2$  are the timelike scalar form factor  and the Gegenbauer coefficient respectively. As a first approximation, the S-wave resonances are used to parametrized $F_s(m_{\pi\pi}^2)$, to include both resonant and nonresonant contributions into the S-wave two-pion distribution amplitudes. Therefore, we take into account $f_0(980),f_0(1500)$ and $f_0(1790)$ for the $s\bar s$ density operator, $f_0(500)$ for the $u\bar u$ density operator:
\begin{eqnarray}
F_s^{s\bar s}(m_{\pi\pi}^2)&=&\frac{c_1m_{f_0(980)}^2 e^{i\theta_1}}{m^2_{f_0(980)}-m_{\pi\pi}^2-i m_{f_0(980)}(g_{\pi\pi}\rho_{\pi\pi}+g_{KK}\rho_{KK})}\nonumber\\
&&+\frac{c_2m_{f_0(1500)}^2 e^{i\theta_2}}{m^2_{f_0(1500)}-m_{\pi\pi}^2-i m_{f_0(1500)}\Gamma_{f_0(1500)}(m_{\pi\pi}^2)}\nonumber\\
&&+\frac{c_3m_{f_0(1790)}^2 e^{i\theta_3}}{m^2_{f_0(1790)}-m_{\pi\pi}^2-i m_{f_0(1790)}\Gamma_{f_0(1790)}(m_{\pi\pi}^2)},\nonumber\\
F_s^{u\bar u}(m_{\pi\pi}^2)&=&\frac{c_0 m_{f_0(500)}^2}{m^2_{f_0(500)}-m_{\pi\pi}^2-i m_{f_0(500)}\Gamma_{f_0(500)}(m_{\pi\pi}^2)}.
\end{eqnarray}
$c_0$, $c_i$ and $\theta_i$, $i=1,2,3$, are tunable parameters.   $m_S$ is the pole mass of the resonance, and $\Gamma_S(m_{\pi\pi})$ is the energy-dependent width for a S-wave resonance decaying into two pions. For the contribution of $f_0(980)$,   the Flatt\'{e} model has been used, and the phase space factors $\rho_{\pi\pi}$ and $\rho_{KK}$ are given as~\cite{Aaij:2014emv}
\begin{eqnarray}
\rho_{\pi\pi}=\frac{2}{3}\sqrt{1-\frac{4m_{\pi^\pm}^2}{m_{\pi\pi}^2}}+\frac{2}{3}\sqrt{1-\frac{4m_{\pi^0}^2}{m_{\pi\pi}^2}},
\rho_{KK}=\frac{1}{2}\sqrt{1-\frac{4m_{K^\pm}^2}{m_{\pi\pi}^2}}+\frac{1}{2}\sqrt{1-\frac{4m_{K^0}^2}{m_{\pi\pi}^2}}.
\end{eqnarray}

\section{Perturbative Calculations}
\label{sec:perturbative calculations}
According to factorization theorems, the amplitude for the process can be  calculated as an expansion of $\alpha_s(Q)$ and $\Lambda/Q$, Q denotes a large momentum transfer, and $\Lambda$ is a small hadronic scale. Usually, the factorization formula for the nonleptonic b-meson decays can be expressed as
\begin{eqnarray}
A\sim \int_0^1 dx_1dx_2dx_3 \int d^2\mathbf{b}_1 d^2\mathbf{b}_2 d^2 \mathbf{b}_3\; C(t) \phi_{B}(x_1,\mathbf{b}_1,t) H(x_1,x_2,x_3,\mathbf{b}_1,\mathbf{b}_2,\mathbf{b}_3,t)\phi_{2}(x_2,\mathbf{b}_2,t) \phi_3(x_3,\mathbf{b}_3,t),
\end{eqnarray}
where the Wilson coefficients $C(t)$, organizing the large logarithms from the hard gluon corrections, is described by the renormalization-group summation of QCD dynamics between W boson $m_W$ and the typical scale $t$. The hard kernel $H(x_i,\mathbf{b}_i,t)$, representing $b$-quark decay sub-amplitude, and the nonperturbative meson wave function $\phi_{i}(x_i,\mathbf{b}_i,t)$, describes the evolution from scale $t$ to the lower hadronic scale $\Lambda_{QCD}$. For a review of this approach, see Ref.~\cite{Li:2003yj}.

\begin{figure}
  \centering
  \includegraphics[width=0.8\columnwidth]{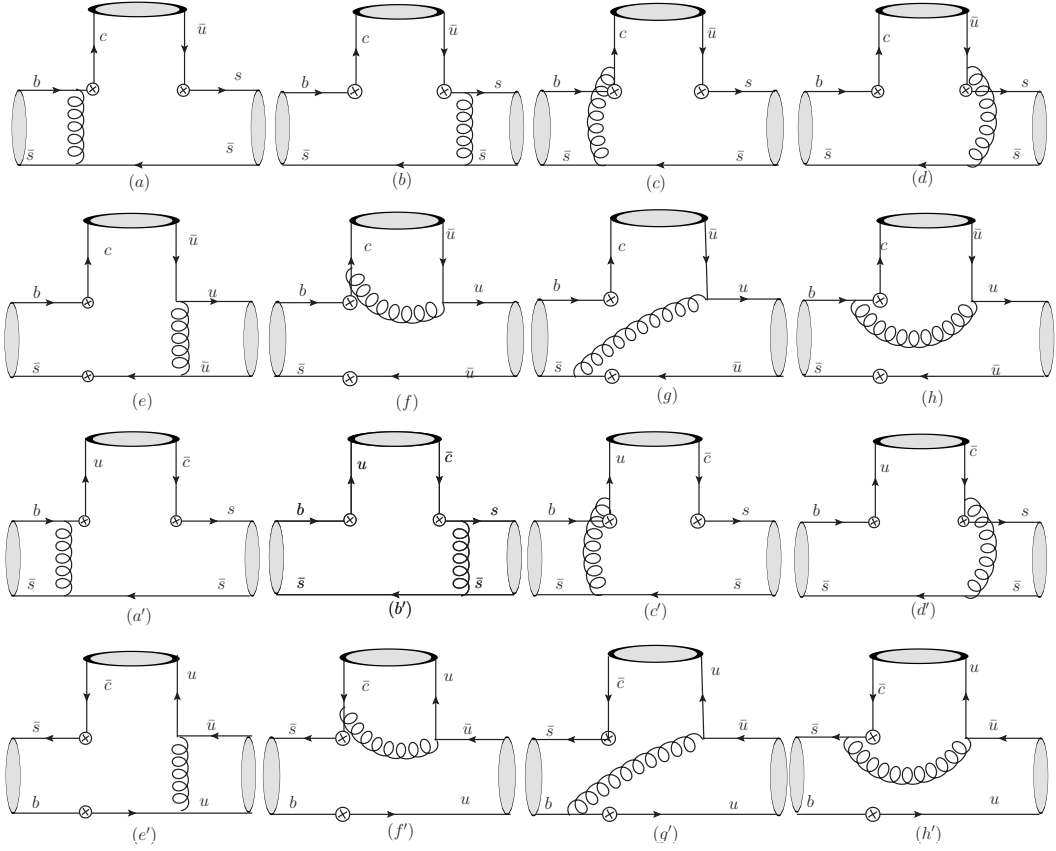}\\
  \caption{Typical Feynman diagrams for the three-body decays $\bar B_s^0 \to D^0(\bar D^0)\pi^+\pi^-$. For the three-body process, the operators in quark level are $\mathcal{O}_1, \mathcal{O}_2$, which correspond with two kinds of Feynman diagrams: the color-suppressed and the W-exchange. The color-suppressed diagrams are drawn in panels ($a$-$d$) and ($a'$-$d'$), further more, the W-exchange diagrams are shown in ($e$-$h$) and ($e'$-$h'$).}\label{fig:fig1}
\end{figure}

The effective Hamiltonian for $\bar B_s^0\to D^0(\bar D^0) \pi^+\pi^-$ is given as
\begin{eqnarray}
\mathcal{H}_{eff}=\frac{G_F}{\sqrt{2}} V_{Qb}V_{qs} (C_1  {O}_1+C_2  {O}_2), \; (Q=c,u,\ q=u,c).
\end{eqnarray}
with $ {O}_1=(\bar c_{\alpha} b_{\beta})_{V-A}(\bar s_{\beta} u_{\alpha})_{V-A}$, $ {O}_2=(\bar c_{\alpha} b_{\alpha})_{V-A}(\bar s_{\beta} u_{\beta})_{V-A}$ for the $\bar B_s^0 \to D^0 \pi^+\pi^-$ process, and $ {O}_1=(\bar u_{\alpha} b_{\beta})_{V-A}(\bar s_{\beta} c_{\alpha})_{V-A}$, $ {O}_2=(\bar u_{\alpha} b_{\alpha})_{V-A}(\bar s_{\beta} c_{\beta})_{V-A}$ for the process of $\bar B_s^0 \to \bar D^0 \pi^+\pi^-$. In particular, the penguin operators do not contribute  to the processes. Using  the above effective Hamiltonian, we obtain the typical Feynman diagrams for the $\bar B_s^0 \to D^0 \pi^+\pi^-$ process shown in Fig.~\ref{fig:fig1}, in which the first row represents the color-suppressed emission process, and the second row indicates the W-exchange process. In the factorization framework, the factorizable diagrams in Fig.~\ref{fig:fig1}(a,b,e,f) are relevant to $a_2$, and the non-factorizable diagrams in Fig.~\ref{fig:fig1}(c,d,g,h) are proportional to $C_2$~\cite{Buchalla:1995vs}, where
\begin{eqnarray}
a_1=C_2+C_1/N_c,\; a_2=C_1+C_2/N_c.
\end{eqnarray}

We will work in the light-cone coordinates. The momentum of the mesons are defined as follows:
\begin{eqnarray}
P_{B_s}=(p_1^+,p_1^-,0_\perp),\; P_{\pi\pi}=(p_2^+,0,0_\perp),\; P_D=(p_1^+-p_2^+,m_{B_s}^2/(2p_1^+),0_\perp).
\end{eqnarray}
Accordingly, the transfer momentum and light-cone components can be achieved as $q^2=(P_{B_s}-P_{\pi\pi})^2=(1-\rho)m_{B_s}^2$, $\rho=1-\frac{m_D}{m_{B_s}}$, $p_1^-=m_{B_s}^2/(2p_1^+)$ and $p_2^+=(m_{B_s}^2-q^2)p_1^+/m_{B_s}^2$. In the heavy quark limit, the mass difference of b-quark(c-quark) and $B_s$(D) meson is negligible, $m_{B_s,D}=m_{b,c} +\bar \Lambda$($\bar \Lambda$ is the order of QCD scale). Since $m_{B_s}\gg m_{D}\gg \bar \Lambda$, we  expand the amplitudes in terms of  $\frac{m_D}{m_{B_s}}$, $\frac{\bar \Lambda}{m_D}$ and high order $\frac{\bar \Lambda}{m_{B_s}}$. At the leading order of expansion, $\rho\sim1, q^2\sim0$. The momenta  of the light quark in mesons ($k_1,k_3$ represent the momentum of light quark in $B_s$ and $D$ meson, $k_2$ is the momentum of positive quark in pion-pair system) are given as
\begin{eqnarray}
k_1=(0,x_1 P_{B_s}^-,k_{1\perp}),\; k_2=(x_2 P_{\pi\pi}^+,0,k_{2\perp}),\; k_3=(0,x_3 P_D^-,k_{3\perp}).
\end{eqnarray}
In the $k_T$-factorization, the color-suppressed emission Feynman diagrams can be calculated out, with the formulas labelling  as $e_{x}$(x=1,2,3,4) in subscript. Thus factorization formulas for the color-suppressed $D^0$-emission diagrams are given as
\begin{eqnarray}
\mathcal{M}_{e12}&=&8\pi C_F m_{B_s}^4 f_D\int^1_0dx_1dx_2\int^{1/\Lambda}_0b_1db_1b_3db_2\phi_B(x_1,b_1)\{E_{e_1}(t_{e_1})h_{e_1}(x_1,x_2,b_1,b_2)a_2(t_{e_1})\nonumber\\
&&[r_0 (1-2x_2)(\phi^s_{\pi\pi}(s\bar s,x_2)-\phi^T_{s\bar s,\pi\pi}(x_2))\notag
+(2-x_2)\phi_{\pi\pi}(s\bar s,x_2)]-2r_0 \phi^s_{\pi\pi}(s\bar s,x_2)E_{e_2}(t_{e_2})h_{e_2}(x_1,x_2,b_1,b_2)a_2(t_{e_2})\},\notag\nonumber\\
\mathcal{M}_{e34}&=&\frac{32\pi C_F m_{B_s}^4 }{\sqrt{2N_c}} \int^1_0dx_1d_2dx_3\int^{1/\Lambda}_0b_1db_1b_3db_3\phi_B(x_1,b_1)\phi_D(\bar x_3,b_3)C_2(t_{e_3})\nonumber\\
&&\{E_{e_3}(t_{e_3})h_{e_3}(x_1,x_2,x_3,b_1,b_3)
[r_0 \bar x_2 (\phi^s_{\pi\pi}(s\bar s,x_2)+\phi^T_{\pi\pi}(s\bar s,x_2))\notag
+x_3\phi_{\pi\pi}(s\bar s,x_2)]\nonumber\\
&&-E_{e_4}(t_{e_4})h_{e_4}(x_1,x_2,x_3,b_1,b_3)[r_0 \bar x_2(\phi^s_{\pi\pi}(s\bar s,x_2)-\phi^T_{\pi\pi}(s\bar s,x_2))+(\bar x_3+\bar x_2)\phi_{\pi\pi}(s\bar s,x_2)]\},
\end{eqnarray}
where $r_0=\frac{m_{\pi\pi}}{m_{B_s}}$, $C_F$ is the color factor. $\phi_{\pi\pi}(s\bar s,x_2)$ represents the two-pion distribution amplitude defined by $s\bar s$ operator. The hard kernels $E_{e_x}$ and $h_{e_x}$   are given in the following.

The factorization formulas for the W-exchange $D^0$ diagrams $\mathcal{M}_{w12}$ and  $\mathcal{M}_{w34}$  are given   as
\begin{eqnarray}
\mathcal{M}_{w12}&=&8\pi C_F m_{B_s}^4 f_{B_s}\int^1_0dx_2dx_3\int^{1/\Lambda}_0b_2db_2b_3db_3\phi_D(x_3,b_3)\{E_{w_1}(t_{w_1})h_{w_1}(x_2,x_3,b_2,b_3)a_2(t_{w_1})\nonumber\\
&&[x_3 \phi_{\pi\pi}(u\bar u,x_2)+2 r_0 r_D (x_3+1)\phi^s_{\pi\pi}(u\bar u,x_2)\notag]-[x_2\phi_{\pi\pi}(u\bar u,x_2)-r_0r_D(2x_2+1)
\phi^s_{\pi\pi}(u\bar u,x_2)\nonumber\\&&+r_0r_D(1-2x_2)\phi^T_{\pi\pi}(u\bar u,x_2)]E_{w_2}(t_{w_2})h_{w_2}(x_2,x_3,b_2,b_3)a_2(t_{w_2})\},\nonumber\\
\mathcal{M}_{w34}&=&\frac{32\pi C_F m_{B_s}^4 }{\sqrt{2N_c}}\int^1_0dx_1dx_2dx_3\int^{1/\Lambda}_0b_1db_1b_2db_2\phi_{B_s}(x_1,b_1)\phi_D(x_3,b_2)\{E_{w_3}(t_{w_3})h_{w_3}(x_1,x_2,x_3,b_1,b_2)C_2(t_{w_3})\nonumber\\
&&[x_2 \phi_{\pi\pi}(u\bar u,x_2)+ r_0 r_D (x_2+x_3)\phi^s_{\pi\pi}(u\bar u,x_2)+r_0r_D(x_2-x_3)\phi^T_{\pi\pi}(u\bar u,x_2) \notag]+[-x_3\phi_{\pi\pi}(u\bar u,x_2)\nonumber\\&&-r_0r_D(x_2+x_3+2)
\phi^s_{\pi\pi}(u\bar u,x_2)+r_0r_D(x_2-x_3)\phi^T_{\pi\pi}(u\bar u,x_2)]E_{w_4}(t_{w_4})h_{w_4}(x_1,x_2,x_3,b_1,b_2)C_2(t_{w_4})\},
\end{eqnarray}
where $r_D=\frac{m_{D}}{m_{B_s}}$, $\phi_{\pi\pi}(u\bar u,x_2)$ represents the distribution amplitude of the $u\bar u$ operator. Due to the helicity  suppression, the contribution of factorizable diagrams $\mathcal{M}_{w12}$ is suppressed significantly. Therefore, the dominant contribution comes from the non-factorizable diagrams $\mathcal{M}_{w34}$.

In the $\bar D^0$-emission process, the two factorizable diagrams  have the same factorization $\mathcal{M}_{e12}=\mathcal{M}_{e'12}$. Accordingly, we give the factorization  formulas for the nonfactorizable emission diagrams $\mathcal{M}_{e'34}$,  the factorizable W-exchange diagrams $\mathcal{M}_{w'12}$ and   the nonfactorizable W-exchange diagrams $\mathcal{M}_{w'34}$ as follows:
\begin{eqnarray}
\mathcal{M}_{e'34}&=&\frac{32\pi C_F m_{B_s}^4 }{\sqrt{2N_c}} \int^1_0dx_1d_2dx_3\int^{1/\Lambda}_0b_1db_1b_3db_3\phi_B(x_1,b_1)\phi_D(\bar x_3,b_3)\nonumber\\
&&\{E_{e'_3}(t_{e'_3})h_{e'_3}(x_1,x_2,x_3,b_1,b_3)C_2(t_{e'_3})[r_0 (\bar x_2)(\phi^s_{\pi\pi}(s\bar s,x_2)+\phi^T_{\pi\pi}(s\bar s,x_2))\notag
+x_3\phi_{\pi\pi}(s\bar s,x_2)]\nonumber\\
&&-E_{e'_4}(t_{e'_4})h_{e'_4}(x_1,x_2,x_3,b_1,b_3)C_2(t_{e'_4})[r_0 \bar x_2(\phi^s_{\pi\pi}(s\bar s,x_2)-\phi^T_{\pi\pi}(s\bar s,x_2))+(\bar x_3+\bar x_2)\phi_{\pi\pi}(s\bar s,x_2)]\},\nonumber\\
\mathcal{M}_{w'12}&=&8\pi C_F m_{B_s}^4 f_{B_s}\int^1_0dx_2dx_3\int^{1/\Lambda}_0b_2db_2b_3db_3\phi_{\bar D}(x_3,b_3)\{E_{w'_1}(t_{w'_1})h_{w'_1}(x_2,x_3,b_2,b_3)a_2(t_{w'_1})\nonumber\\
&&[(1-x_2) \phi_{\pi\pi}(u\bar u,x_2)+ r_0 r_D (2x_2-3)\phi^s_{\pi\pi}(u\bar u,x_2)+r_0r_D (1-2x_2)\phi^T_{\pi\pi}(u\bar u,x_2)\notag]\nonumber\\&&+[-x_3\phi_{\pi\pi}(u\bar u,x_2)+2r_0r_D(x_3+1)
\phi^s_{\pi\pi}(u\bar u,x_2)]E_{w'_2}(t_{w'_2})h_{w'_2}(x_2,x_3,b_2,b_3)a_2(t_{w'_2})\},\notag\nonumber\\
\mathcal{M}_{w'34}&=&\frac{32\pi C_F m_{B_s}^4 }{\sqrt{2N_c}}\int^1_0dx_1dx_2dx_3\int^{1/\Lambda}_0b_1db_1b_2db_2\phi_{B_s}(x_1,b_1)\phi_{\bar D}(x_3,b_2)\{E_{w'_3}(t_{w'_3})h_{w'_3}(x_1,x_2,x_3,b_1,b_2)C_2(t_{w'_3})\nonumber\\
&&[x_3 \phi_{\pi\pi}(u\bar u,x_2)- r_0 r_D (1-x_2+x_3)\phi^s_{\pi\pi}(u\bar u,x_2)+r_0r_D(x_2+x_3-1)\phi^T_{\pi\pi}(u\bar u,x_2) \notag]+[(x_2-1)\phi_{\pi\pi}(u\bar u,x_2)\nonumber\\&&+r_0r_D(-x_2+x_3+3)
\phi^s_{\pi\pi}(u\bar u,x_2)+r_0r_D(x_2+x_3-1)\phi^T_{\pi\pi}(u\bar u,x_2)]E_{w'_4}(t_{w'_4})h_{w'_4}(x_1,x_2,x_3,b_1,b_2)C_2(t_{w'_4})\}.\nonumber\\
\end{eqnarray}

In the following, we give the forms for the offshellness of the intermediate gluon $\beta_{e_x}$/$\beta_{w_x}$ and quarks $\alpha_{e_x}$/$\alpha_{w_x}$($x=1,2,3,4$) in the $\bar B_s^0 \to D^0 \pi^+\pi^-$ process.
\begin{eqnarray}
&&\alpha_{e_1}=(1-x_2) m_{B_s}^2 \rho,\; \alpha_{e_2}=x_1 m_{B_s}^2 \rho,\; \alpha_{e_3}=x_1(1-x_2)m_{B_s}^2 \rho,\; \alpha_{e_4}=x_1(1-x_2)m_{B_s}^2 \rho,\nonumber\\
&&\alpha_{w_1}=x_3 m_{B_s}^2 \rho,\; \alpha_{w_2}=(1-\rho+x_2\rho) m_{B_s}^2,\; \alpha_{w_3}=x_2x_3m_{B_s}^2 \rho,\; \alpha_{w_4}=x_2x_3m_{B_s}^2 \rho,\nonumber\\
&&\beta_{e_1}=x_1(1-x_2)m_{B_s}^2 \rho,\; \beta_{e_2}=x_1(1-x_2)m_{B_s}^2 \rho,\; \nonumber\\ &&\beta_{e_3}=[(x_1-x_3)(1-x_2\rho)+(1-\rho)]m_{B_s}^2,\; \beta_{e_4}=(1-x_2)(x_1+x_3-1)m_{B_s}^2\rho,\nonumber\\
&&\beta_{w_1}=x_2x_3m_{B_s}^2 \rho,\; \beta_{w_2}=x_2x_3m_{B_s}^2 \rho,\; \nonumber\\ &&\beta_{w_3}=(x_3-x_1)x_2 m_{B_s}^2\rho,\; \beta_{w_4}=((1-x_1-x_3)(1-x_2\rho)-1)m_{B_s}^2.
\end{eqnarray}
For the  $B_s^0 \to \bar D^0 \pi^+\pi^-$, we have
\begin{eqnarray}
&&\alpha_{e'_1}=(1-x_2) m_{B_s}^2 \rho,\; \alpha_{e'_2}=x_1 m_{B_s}^2 \rho,\; \alpha_{e'_3}=x_1(1-x_2)m_{B_s}^2 \rho,\; \alpha_{e'_4}=x_1(1-x_2)m_{B_s}^2 \rho,\nonumber\\
&&\alpha_{w'_1}=(1-x_2\rho) m_{B_s}^2,\; \alpha_{w'_2}=x_3 m_{B_s}^2 \rho,\; \alpha_{w'_3}=x_3(1-x_2)m_{B_s}^2 \rho,\; \alpha_{w'_4}=x_3(1-x_2)m_{B_s}^2 \rho,\nonumber\\
&&\beta_{e'_1}=x_1(1-x_2)m_{B_s}^2 \rho,\; \beta_{e'_2}=x_1(1-x_2)m_{B_s}^2 \rho,\; \nonumber\\ &&\beta_{e'_3}=(1-x_2)(x_1-x_3)m_{B_s}^2\rho,\; \beta_{e'_4}=[(x_1+x_3-1)(1-x_2\rho)+(1-\rho)]m_{B_s}^2,\nonumber\\
&&\beta_{w'_1}=x_3(1-x_2)m_{B_s}^2 \rho,\; \beta_{w'_2}=x_3(1-x_2)m_{B_s}^2 \rho,\; \nonumber\\ &&\beta_{w'_3}=(1-x_2)(x_3-x_1)m_{B_s}^2\rho,\; \beta_{w'_4}=((1-x_1-x_3)(1-\rho+x_2\rho)-1)m_{B_s}^2.
\end{eqnarray}
The hard kernel functions $h_{e_x}$($h_{e'_x}$) and $h_{w_x}$($h_{w'_x}$) are written as
\begin{eqnarray}
&&h_{e_i}(x_1,x_2,b_1,b_2)=[\theta(b_1-b_2) I_0(\sqrt{\alpha_{e_i}}b_2) K_0(\sqrt{\beta_{e_i}}b_1)+(b_1\leftrightarrow b_2)] K_0(\sqrt{\beta_{e_i}}b_1)S_t(\alpha_{e_i}/(m_{B_s}^2\rho)),\nonumber\\
&&h_{e_j}(x_1,x_2,x_3,b_1,b_3)=[\theta(b_1-b_3) I_0(\sqrt{\alpha_{e_j}}b_3) K_0(\sqrt{\beta_{e_j}}b_1)+(b_1\leftrightarrow b_3)] \times \begin{cases}K_0(\sqrt{\beta_{e_j}}b_1),& \beta_{e_j}\geq0,\\ \frac{i\pi}{2}H_0^{(1)}(\sqrt{|\beta_{e_j}|}b_1), &\beta_{e_j}<0, \end{cases},\nonumber\\
&&h_{w_k}(x_1,x_2,b_2,b_3)=(i\frac{\pi}{2})^2 H_0^{(1)}(\sqrt{\beta_{w_k}}b_2)[\theta(b_2-b_3)H_0^{(1)}(\sqrt{\alpha_{w_k}}b_2)J_0(\sqrt{\alpha_{w_k}}b_3)+(b_2\leftrightarrow b_3)]S_t(\alpha_{w_k}/(m_{B_s}^2\rho)),\nonumber\\
&&h_{w_l}(x_1,x_2,x_3,b_1,b_2)=i\frac{\pi}{2}\Bigg[\theta(b_1-b_2)H_0^{(1)}(\sqrt{\alpha_{w_l}}b_1)J_0(\sqrt{\alpha_{w_l}}b_2)+(b_1\leftrightarrow b_2)\Bigg]\times \begin{cases}K_0(\sqrt{\beta_{w_l}}b_1),& \beta_{w_l}\leq0,\\ \frac{i\pi}{2}H_0^{(1)}(\sqrt{|\beta_{w_l}|}b_1), &\beta_{w_l}>0, \end{cases}.\nonumber\\
\end{eqnarray}
where $i,k=1,2$ and $j,l=3,4$, the $I_0$, $K_0$ and $H_0=J_0+i Y_0$ are Bessel functions. The threshold resummation factor $S_t(x)$ follows the parametrization as
\begin{eqnarray}
S_t(x)=\frac{2^{1+2c \Gamma(3/2+c)}}{\sqrt{\pi}\Gamma(1+c)}[x(1-x)]^c,
\end{eqnarray}
with the parameter $c=0.4$ in this paper.
The evolution factors $E_{x}(t)$s in the factorization formulas are given by
\begin{eqnarray}
E_{e_i}(t)&=&\alpha_s(t) {\rm exp}(-S_{B_s}(t)-S_{\pi\pi}(t)),\nonumber\\
E_{e_j}(t)&=&\alpha_s(t) {\rm exp}(-S_{B_s}(t)-S_{\pi\pi}(t)-S_{D}(t))|_{b_{1}=b_{2}},\nonumber\\
E_{w_k}(t)&=&\alpha_s(t) {\rm exp}(-S_{\pi\pi}(t)-S_{D}(t)),\nonumber\\
E_{w_l}(t)&=&\alpha_s(t) {\rm exp}(-S_{B_s}(t)-S_{\pi\pi}(t)-S_{D}(t))|_{b_2=b_{3}},\nonumber\\
\end{eqnarray}
where
\begin{eqnarray}
S_{B_s}(t)&=&s(x_1 m_{B_s}, b_1)+\frac{5}{3}\int_{1/b_1}^t \frac{d\bar \mu}{\bar \mu}\gamma_{q}(\alpha_s(\bar \mu)),\nonumber\\
S_{D}(t)&=&s(x_3 m_{B_s}, b_3)+2\int_{1/b_3}^t \frac{d\bar \mu}{\bar \mu}\gamma_{q}(\alpha_s(\bar \mu)),\nonumber\\
S_{\pi\pi}(t)&=&s(x_2 m_{B_s}, b_2)+s((1-x_2) m_{B_s}, b_2)+2\int_{1/b_2}^t \frac{d\bar \mu}{\bar \mu}\gamma_{q}(\alpha_s(\bar \mu)),
\end{eqnarray}
with the quark anomalous dimension $\gamma_{q}=-\alpha_s/\pi$. The explicit expression of $s(Q,b)$ can be found, for example, in Appendix A of Ref~\cite{Ali:2007ff}. The hard scales are chosen as
\begin{eqnarray}
t_{e_i}&=&max(\sqrt{\alpha_{e_i}},\sqrt{\beta_{e_i}},1/b_1,1/b_2),\;
t_{e_j}=max(\sqrt{\alpha_{e_j}},\sqrt{\beta_{e_j}},1/b_1,1/b_3),\nonumber\\
t_{w_k}&=&max(\sqrt{\alpha_{w_k}},\sqrt{\beta_{w_k}},1/b_2,1/b_3),
t_{w_l}=max(\sqrt{\alpha_{w_l}},\sqrt{\beta_{w_l}},1/b_1,1/b_2).
\end{eqnarray}

Therefore, we obtain the total decay amplitudes,
\begin{eqnarray}
\mathcal{A}(\bar B_s\to D^0 \pi^+\pi^-)=\frac{G_F}{\sqrt{2}}V_{cb}V_{us}^*(\mathcal{M}_{e12}+\mathcal{M}_{e34}+\mathcal{M}_{w12}+\mathcal{M}_{w34}) ,\nonumber\\
\mathcal{A}(\bar B_s\to \bar D^0 \pi^+\pi^-)=\frac{G_F}{\sqrt{2}}V_{ub}V_{cs}^*(\mathcal{M}_{e'12}+\mathcal{M}_{e'34}+\mathcal{M}_{w'12}+\mathcal{M}_{w'34}).
\end{eqnarray}

The differential branching ratio for the $\bar B_s^0 \to D^0(\bar D^0) \pi^+\pi^-$ decay follows the formula given as~\cite{Beringer:1900zz,Agashe:2014kda}
\begin{eqnarray}
\frac{d\mathcal{B}}{d{m_{\pi\pi}}}=\tau_{B_s} \frac{m_{\pi\pi}|\overrightarrow{p_1}||\overrightarrow{p_3}|}{4(2\pi)^3m_{B_s}^3}|\mathcal{A}|^2,
\end{eqnarray}
with the $B_s$ meson mean lifetime $\tau_{B_s}$. The kinematic variables $|\overrightarrow{p_1}|$ and $|\overrightarrow{p_3}|$ denote the magnitudes of the $\pi^+$ and $D$ momenta in the center-of-mass frame of the pion pair,
\begin{eqnarray}
|\overrightarrow{p_1}|=\frac{1}{2}\sqrt{m_{\pi\pi}^2-4m_{\pi^\pm}^2},\ |\overrightarrow{p_3}|=\frac{1}{2m_{\pi\pi}}\sqrt{[m_{B_s}^2-(m_{\pi\pi}+m_D)^2][m_{B_s}^2-(m_{\pi\pi}-m_D)^2]}.
\end{eqnarray}
\section{Numerical Results}
\label{sec:numerical results}
We adopt the following inputs(in units of GeV)~\cite{Beringer:1900zz,Agashe:2014kda}
\begin{eqnarray*}
\Lambda_{\bar{MS}}^{f=4}=0.250,\ m_{B_s}=5.367,\ m_{D^0}=1.869,\ m_{\pi^{\pm}}=0.140, m_{\pi^0}=0.135,\ m_{K^\pm}=0.494,\ \\ m_{K^0}=0.498,\ m_b=4.66,\ m_s=0.095,\ \tau_{B_s}=1.512\times10^{-12}s,\ G_F=1.166\times10^{-5},
\end{eqnarray*}
and the CKM matrix elements are taken as:
\begin{eqnarray*}
|V_{us}|=0.2252,\ |V_{ub}|=3.89\times10^{-3},\ |V_{cs}|=0.97345,\ |V_{cb}|=40.6\times10^{-3}.
\end{eqnarray*}
The parameters for the scalar form factor $F_s(m_{\pi\pi}^2)$ are extracted from the LHCb data in the process of $B_s\to J/\psi\pi^+\pi^-$, given as~\cite{Aaij:2014emv,Aaij:2014siy} (mass and widths are given in units of GeV):
\begin{eqnarray*}
&&m(f_0(500))=0.5,\ m(f_0(980))=0.97,\ m(f_0(1500))=1.5,\ m(f_0(1790))=1.81,\\
&&\Gamma(f_0(500))=0.4,\ \Gamma(f_0(1500))=0.12,\ \Gamma(f_0(1790))=0.32, \\
&& g_{\pi\pi}=0.167,\ g_{KK}=3.47 g_{\pi\pi},\\
&&c_0=3.500,\ c_1=0.900,\ c_2=0.106,\ c_3=0.066,\ \\
&&\theta_1=-\frac{\pi}{2},\ \theta_2=\frac{\pi}{4},\ \theta_3=0.
\end{eqnarray*}

We calculate the branching ratios with the different resonances in S-wave pion-pair function shown in Tab~\ref{tab:results}. In this table,  the first uncertainties  are from $\omega_b=0.50\pm0.05$ in the $B_s$ wave function, the second errors arise   from $a_2=0.2\pm0.2$ in the pion-pair wave function, and the third uncertainties  come from QCD scale $\Lambda=0.25\pm0.05$. The errors from the parameter of $D$-meson function $C_D$, the variations of CKM matrix elements and the mean lifetime of $B_s$ are tiny, and have been omitted. However the above results are sensitive to $\omega_b$ and $a_2$, namely  the $B_s$ and S-wave two-pion wave functions. The future  measurements of decay branching fractions will be valuable  to understand the $B_s$ physics and the S-wave two-pion resonances.

\begin{table}
\caption{Branching ratios from  the different  intermediate resonances. }\label{tab:results}
\begin{tabular}{c c}
\hline\hline
$Resonances$ & Branching ratio ($\times10^{-6}$)\\
\hline
$\bar B_s^0\to D^0f_0(500)[f_0(500)\to\pi^+\pi^-]$ & $(0.14)_{-0.04}^{+0.05}(\omega_b)_{-0.10}^{+0.22}(a_2)_{-0.01}^{+0.05}(\Lambda_{QCD})$ \\
$\bar B_s^0\to D^0f_0(980)[f_0(980)\to\pi^+\pi^-]$&$(0.52)_{-0.13}^{+0.14}(\omega_b)_{-0.14}^{+0.60}(a_2)_{-0.12}^{+0.10}(\Lambda_{QCD})$ \\
$\bar B_s^0\to D^0f_0(1500)[f_0(1500)\to\pi^+\pi^-]$&$(0.13)_{-0.04}^{+0.04}(\omega_b)_{-0.02}^{+0.09}(a_2)_{-0.03}^{+0.03}(\Lambda_{QCD})$ \\
$\bar B_s^0\to D^0f_0(1790)[f_0(1790)\to\pi^+\pi^-]$&$(0.039)_{-0.011}^{+0.013}(\omega_b)_{-0.003}^{+0.019}(a_2)_{-0.009}^{+0.008}(\Lambda_{QCD})$ \\
$\bar B_s^0\to \bar D^0f_0(500)[f_0(500)\to\pi^+\pi^-]$&$(0.13)_{-0.05}^{+0.05}(\omega_b)_{-0.10}^{+0.23}(a_2)_{-0.03}^{+-0.00}(\Lambda_{QCD})$ \\
$\bar B_s^0\to \bar D^0f_0(980)[f_0(980)\to\pi^+\pi^-]$&$(0.19)_{-0.06}^{+0.07}(\omega_b)_{-0.12}^{+0.19}(a_2)_{-0.01}^{+0.01}(\Lambda_{QCD})$ \\
$\bar B_s^0\to \bar D^0f_0(1500)[f_0(1500)\to\pi^+\pi^-]$&$(0.044)_{-0.014}^{+0.016}(\omega_b)_{-0.025}^{+0.035}(a_2)_{-0.001}^{+0.002}(\Lambda_{QCD})$ \\
$\bar B_s^0\to \bar D^0f_0(1790)[f_0(1790)\to\pi^+\pi^-]$&$(0.013)_{-0.004}^{+0.005}(\omega_b)_{-0.007}^{+0.009}(a_2)_{-0.000}^{+0.000}(\Lambda_{QCD})$\\
\hline
\end{tabular}
\end{table}

Including all the S-wave resonances $f_0(500)$, $f_0(980)$, $f_0(1500)$ and $f_0(1790)$ in the scalar form factor, we obtain the total branching ratio
\begin{eqnarray}
\mathcal{B}(\bar B_s^0\to D^0 (\pi^+\pi^-)_S)=(0.87)^{+0.22}_{-0.20}(\omega_b)^{+1.20}_{-0.32}(a_2)^{+0.13}_{-0.14}(\Lambda_{QCD})\times10^{-6}, \nonumber\\
\mathcal{B}(\bar B_s^0\to \bar D^0 (\pi^+\pi^-)_S)=(0.53)^{+0.20}_{-0.18}(\omega_b)^{+0.66}_{-0.38}(a_2)^{+0.01}_{-0.06}(\Lambda_{QCD})\times10^{-6}.
\end{eqnarray}
We found the $\bar B_s^0\to D^0f_0(500) [f_0(500)\to\pi^+\pi^-]$, $\bar B_s^0\to D^0f_0(980) [f_0(980)\to\pi^+\pi^-]$, $\bar B_s^0\to D^0f_0(1500) [f_0(1500)\to\pi^+\pi^-]$ and $\bar B_s^0\to D^0f_0(1790) [f_0(1790)\to\pi^+\pi^-]$ contributions to be 16.4\%, 59.3\%
, 14.6\% 
 and 4.5\% 
 of the total $\bar B_s^0\to D^0 (\pi^+\pi^-)_S$ decay rate. For the $\bar B_s^0\to \bar D^0 (\pi^+\pi^-)_S$ process, the corresponding rates are 24.6\%, 35.2\%, 8.3\% and 2.4\% respectively. It indicates that the $f_0(500)$ and $f_0(980)$ contributions  are dominant,  and the contribution from $f_0(980)$ is larger than $f_0(500)$ in $D^0$($\bar D^0$) final state. LHCb collaboration measures the branching ratio with the upper limit of $\mathcal{B}(B_s \to \bar D^0 f_0(980))<3.1\times10^{-6}$, which roughly agrees with our value. 

For the comparison of $\bar B_s\to D^0 (\pi\pi)_S$ and $\bar B_s\to \bar D^0 (\pi\pi)_S$, we determine the rate of their branching ratios
\begin{eqnarray}
R_1=\frac{\mathcal{B}(\bar B_s^0\to D^0 (\pi^+\pi^-)_S)}{\mathcal{B}(\bar B_s^0\to \bar D^0 (\pi^+\pi^-)_S)}\sim1.64,
\end{eqnarray}
with the quite different CKM ratio factor
\begin{eqnarray}
R_{CKM}=\Bigg|\frac{V_{cb}V_{us}^*}{V_{ub}V_{cs}^*}\Bigg| \sim5.83.
\end{eqnarray}

\begin{figure}[htbp]
  \centering
  \subfigure[]{
  \begin{minipage}[t]{0.4\linewidth}
  \centering
  \includegraphics[width=0.7\columnwidth]{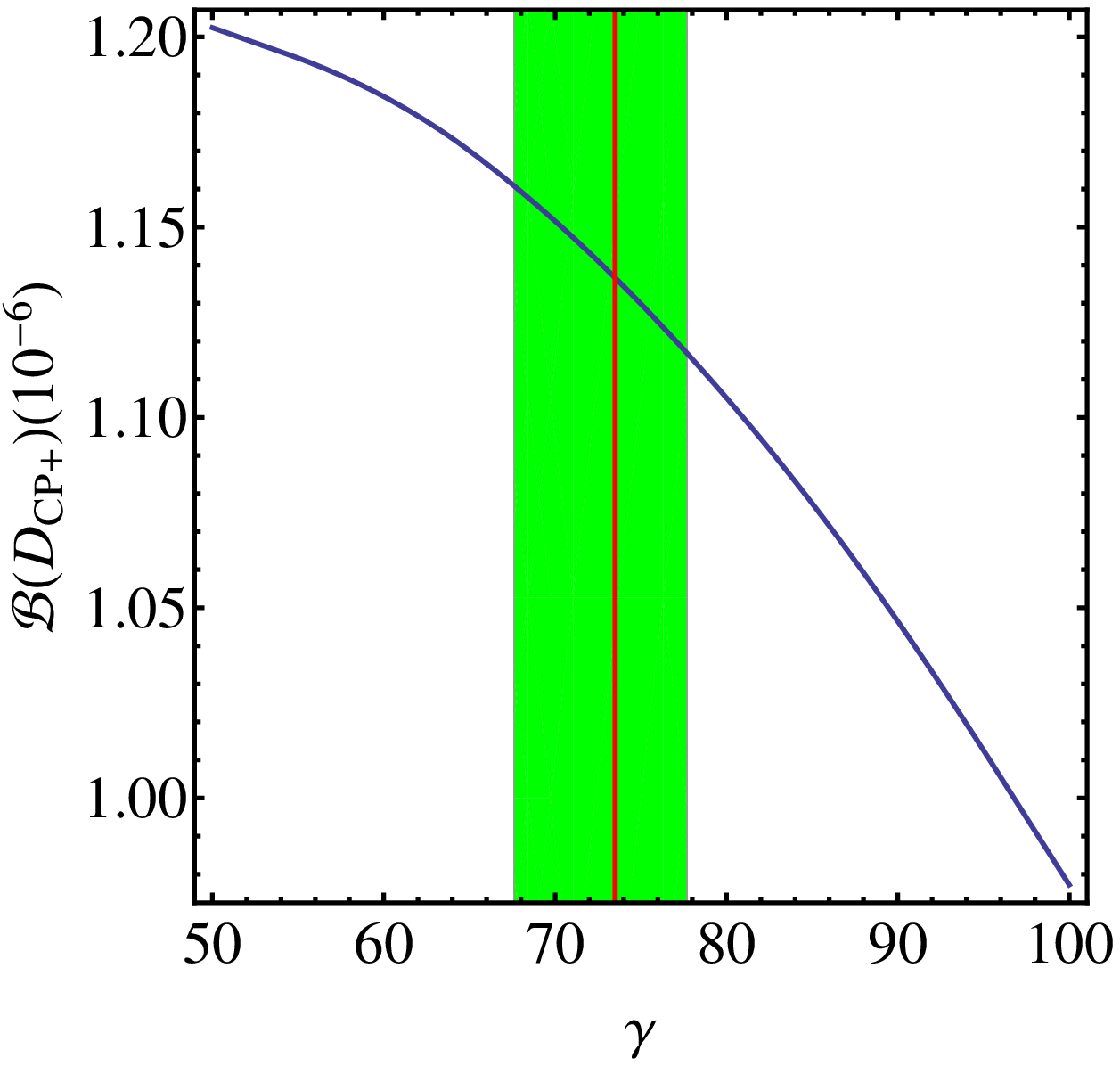}
  \end{minipage}}
  \subfigure[]{
  \begin{minipage}[t]{0.4\linewidth}
  \centering
  \includegraphics[width=0.7\columnwidth]{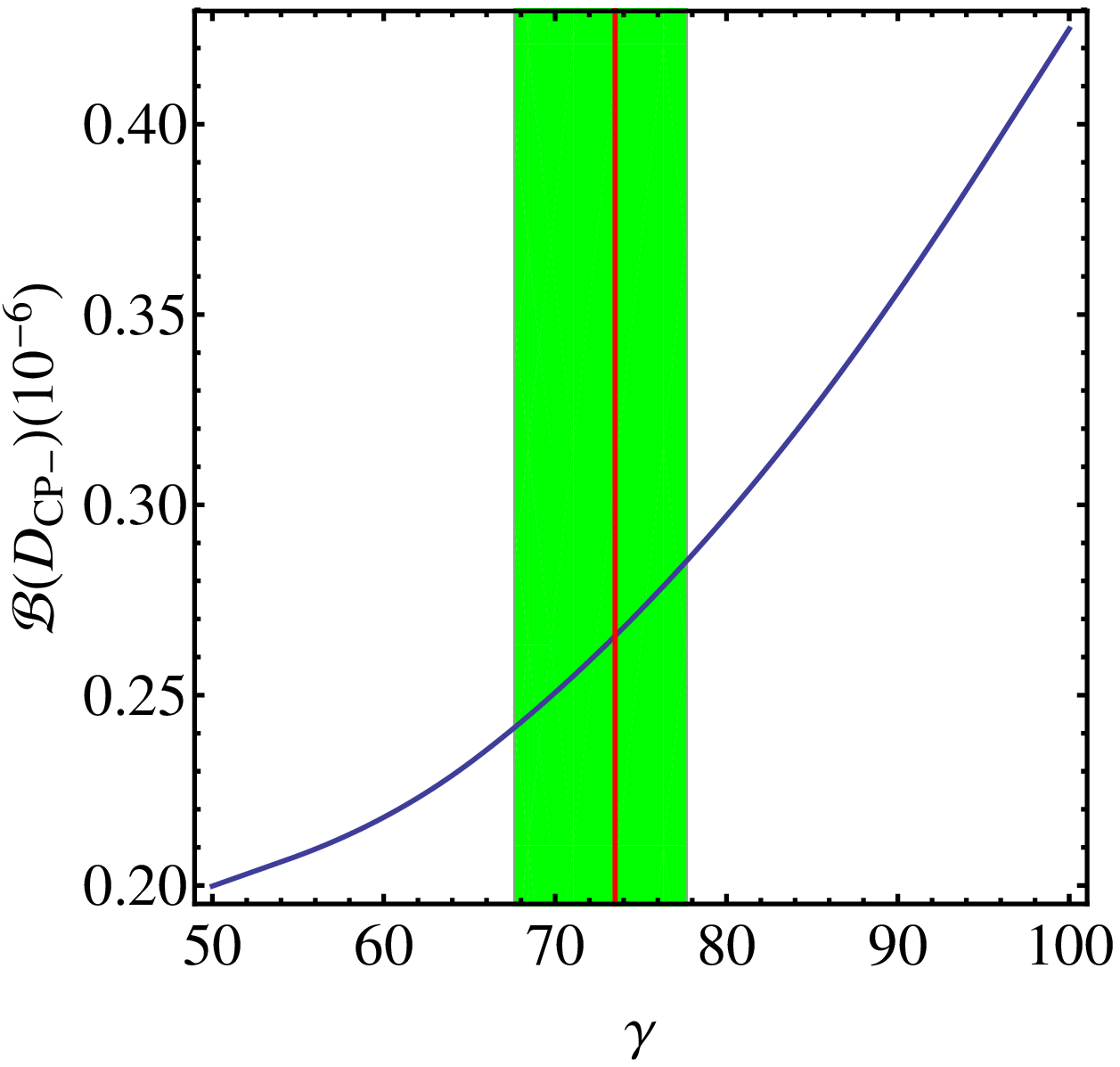}
  \end{minipage}}
    \subfigure[]{
  \begin{minipage}[t]{0.4\linewidth}
  \centering
  \includegraphics[width=0.7\columnwidth]{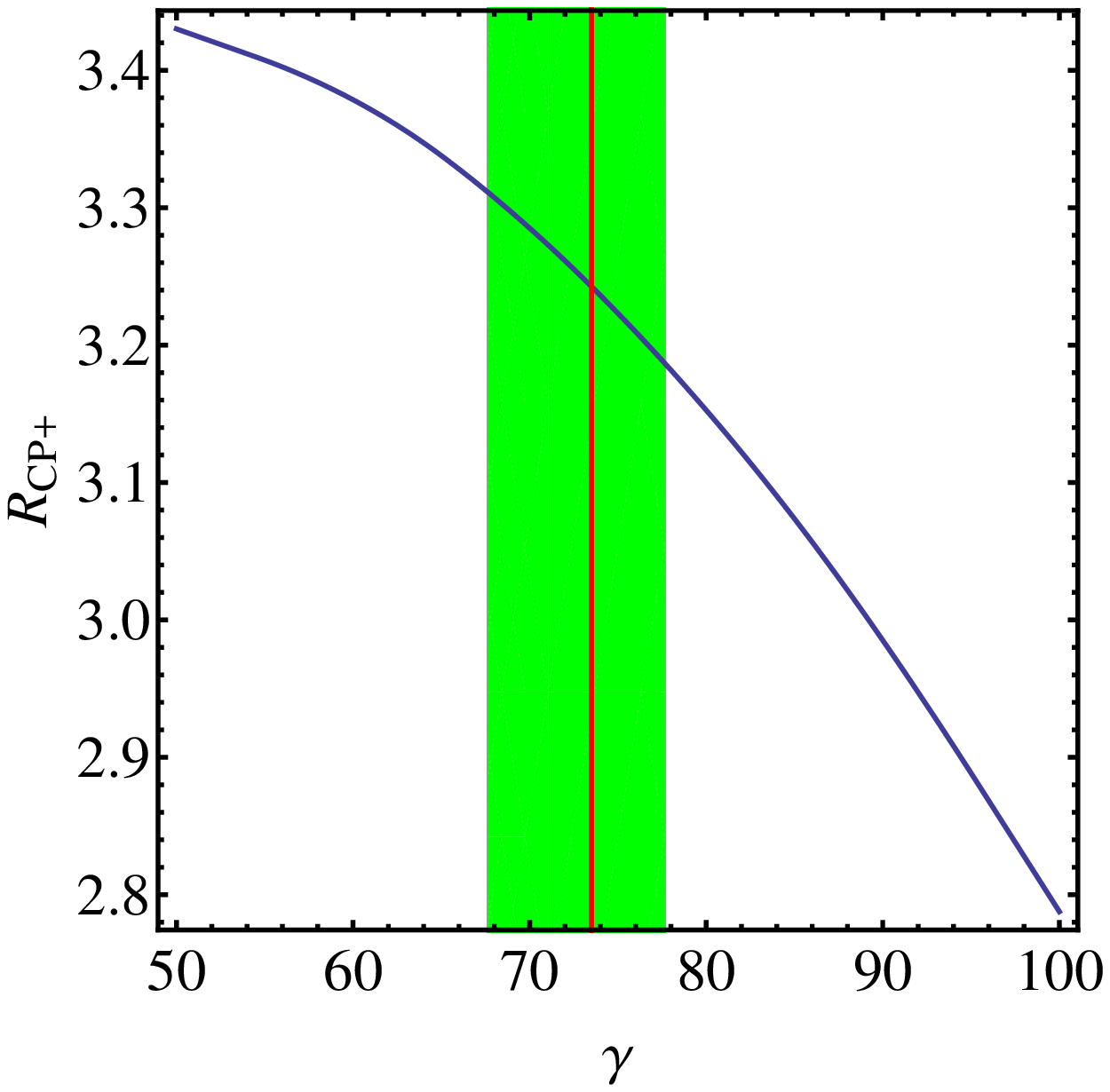}
  \end{minipage}}
    \subfigure[]{
  \begin{minipage}[t]{0.4\linewidth}
  \centering
  \includegraphics[width=0.7\columnwidth]{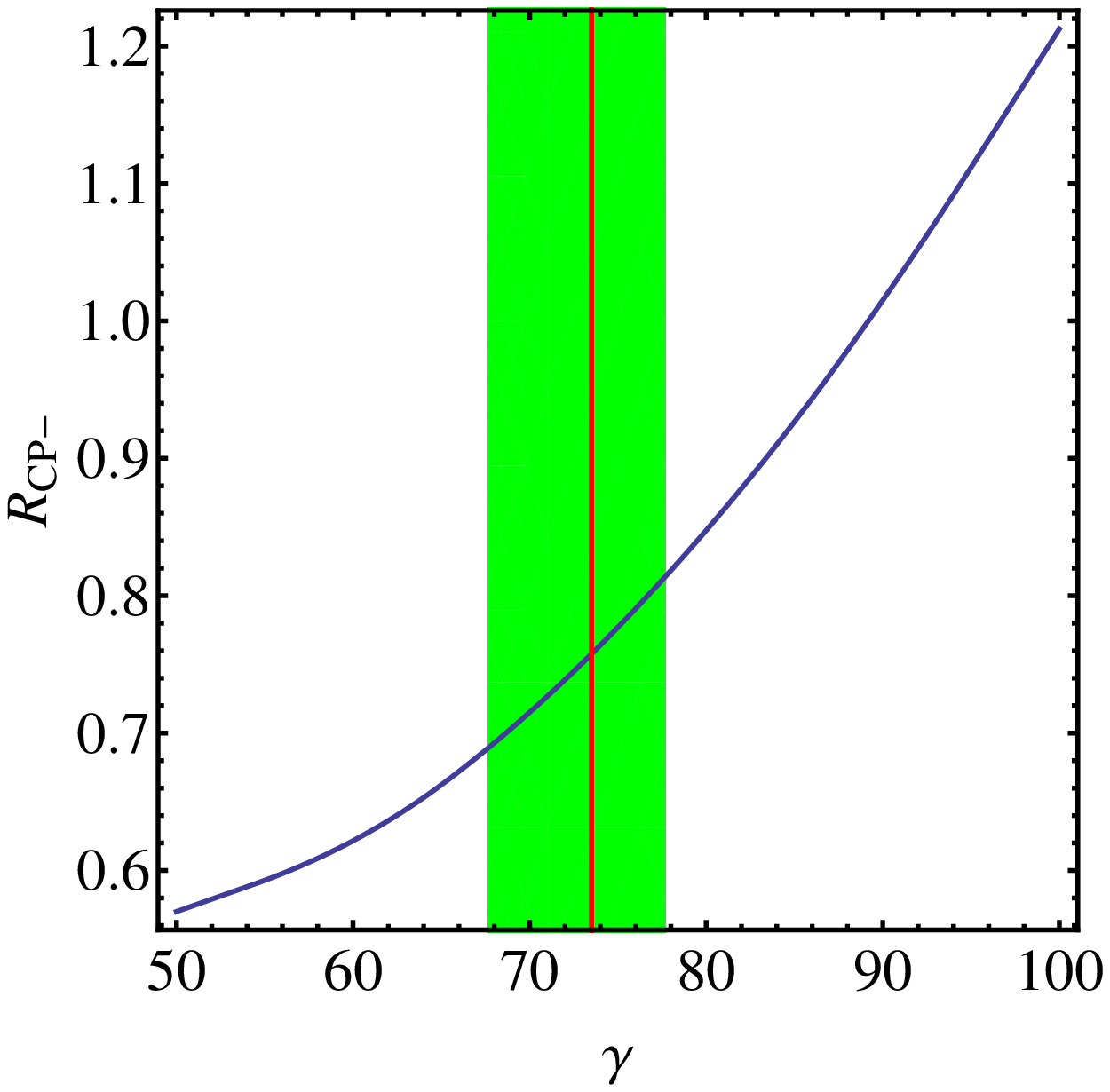}
  \end{minipage}}
  \caption{The dependencies of differential branching ratios $\mathcal{B}(\bar B_s^0 \to D_{CP\pm}(\pi^+\pi^-)_S)$ on $\gamma$ are shown in panels (a,b). For the panels (c,d), the corresponding physical observable measurements $R_{CP\pm}$ are depend on $\gamma$. The shadowed (green) region denotes the current bounds on $\gamma=73.5^{+4.2}_{-5.9}$. }\label{fig:fig3}
\end{figure}

The CKM elements of $\bar B_s^0\to D^0(\bar D^0) (\pi^+\pi^-)_S$ is $V_{cb}V_{us}^*$($V_{ub}V_{cs}^*$), in which $V_{ub}$ is sensitive to the $\gamma$. Therefore, we can achieve the dependence of our results about $\gamma$, by providing a parameter $D_{CP\pm}$ defined as~\cite{Wang:2011zw}
\begin{eqnarray}
\sqrt{2}\mathcal{A}(\bar B_s^0 \to D_{CP\pm}(\pi^+\pi^-)_S)=\mathcal{A}(\bar B_s^0 \to D^0(\pi^+\pi^-)_S )\pm\mathcal{A}(\bar B_s^0 \to \bar D^0(\pi^+\pi^-)_S ).
\end{eqnarray}
Accordingly, the dependence curve of branching ratio $\mathcal{B}(\bar B_s^0 \to D_{CP\pm}(\pi^+\pi^-)_S)$ on $\gamma$ is obtained in Fig.~\ref{fig:fig3}(a,b). In experimentally side, the corresponding physical observable measurement is defined as
\begin{eqnarray}
R_{CP\pm}=\frac{4\mathcal{B}(\bar B_s^0\to D_{CP\pm} (\pi^+\pi^-)_S)}{\mathcal{B}(\bar B_s^0\to D^0 (\pi^+\pi^-)_S)+\mathcal{B}(\bar B_s^0\to \bar D^0 (\pi^+\pi^-)_S)}.
\end{eqnarray}
We give the dependencies of $R_{CP\pm}$ on $\gamma$ shown in Fig.~\ref{fig:fig3}(c,d). The current bound on $\gamma$ is constrained as $\gamma=(73.5^{+4.2}_{-5.9})^\circ$~\cite{Tanabashi:2018oca}.

\begin{figure}[htbp]
  \centering
  \subfigure[]{
  \begin{minipage}[t]{0.45\linewidth}
  \centering
  \includegraphics[width=0.82\columnwidth]{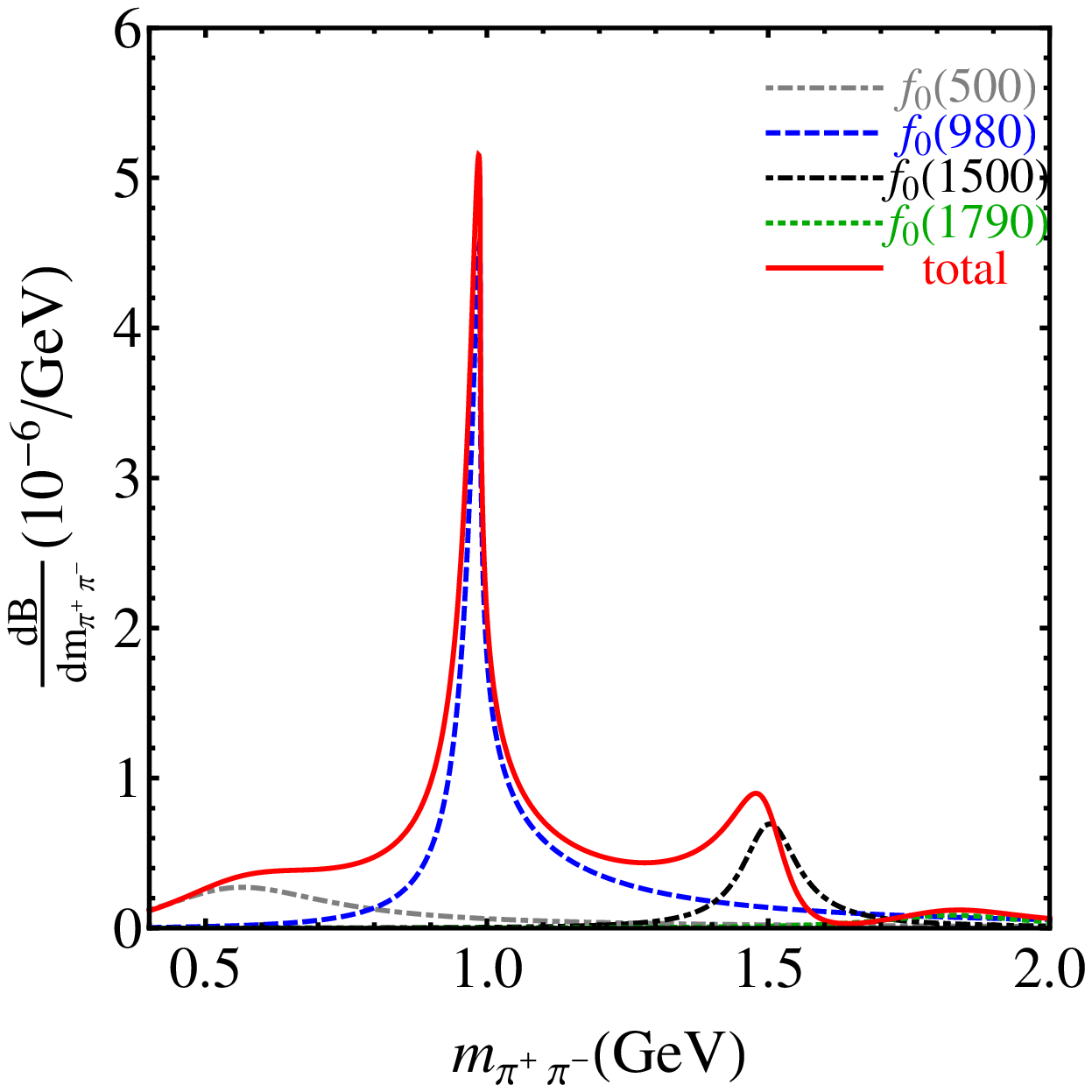}
  \end{minipage}}
  \subfigure[]{
  \begin{minipage}[t]{0.45\linewidth}
  \centering
  \includegraphics[width=0.82\columnwidth]{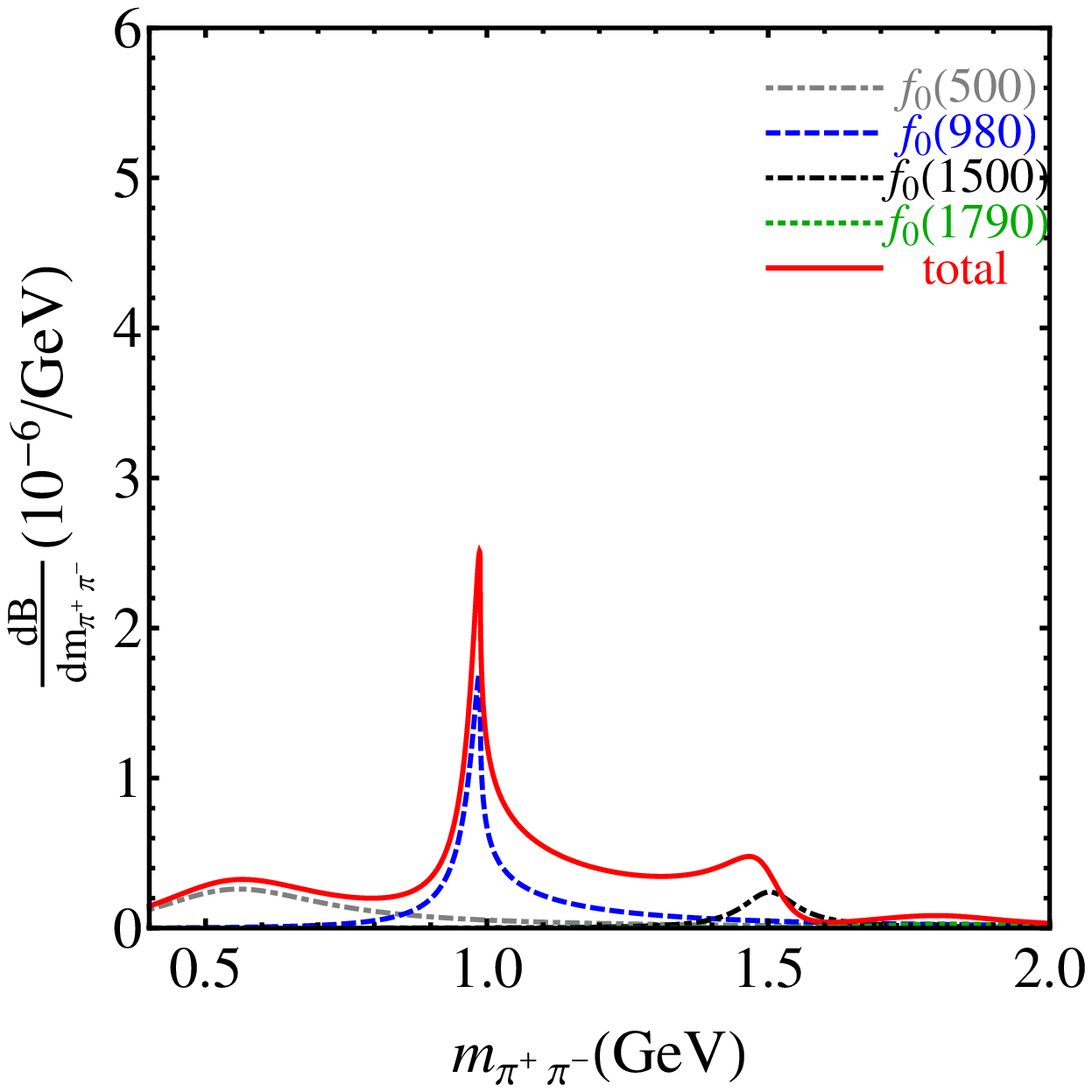}
  \end{minipage}}
  \caption{The differential branching ratios on the pion-pair invariant mass for the resonance $f_0(980)$, $f_0(1500)$ and $f_0(1790)$ in the (a) $\bar B_s^0\to D^0\pi^+\pi^-$ and (b) $\bar B_s^0\to \bar D^0\pi^+\pi^-$ decays.}\label{fig:fig2}
\end{figure}

The predicted dependencies of the differential branching ratios $d\mathcal{B}/dm_{\pi\pi}$ on the pion-pair invariant mass $m_{\pi\pi}$ are presented  in  Fig.~\ref{fig:fig2}.(a) and Fig.~\ref{fig:fig2}.(b)  for the resonances $f_0(500)$, $f_0(980)$, $f_0(1500)$ and $f_0(1790)$ in the $\bar B_s\to D^0 \pi^+\pi^-$ and $\bar B_s\to \bar D^0 \pi^+\pi^-$ decay. The graphs show that the main contribution of the two decays lies in the region around the pole mass $m_{f_0(980)}=0.97$, while the $f_0(500)$ lead to the primary contribution below the region $m_{\pi\pi}=1GeV$. The other resonances $f_0(1500)$ and $f_0(1790)$ still give the considerable contributions to the processes. Therefore, we expect that more precise data from the LHCb and the future KEKB may test our theoretical calculations.



\section{Conclusions}
\label{sec:conclusions}

In the past decades, two-body $B$ decays have provided an ideal platform to extract the standard model parameters, and probe the new physics beyond the SM~\cite{Wang:2014sba,Cerri:2018ypt}.
In this work, we have  studied  the three-body $\bar B_s^0\to D^0(\bar D^0)\pi^+\pi^-$ decay within the PQCD framework, and in particular the S-wave contribution is explicitly calculated.  The   S-wave two-pion light-cone distribution amplitudes can receives both resonant $f_0(500)$, $f_0(980), f_0(1500), f_0(1790)$ and nonresonant contributions. Furthermore, the processes proceed via the tree level operators, and branching ratios are found   in the range from $10^{-7}$ to $10^{-6}$. It is found  that the branching ratios are sensitive to the parameters $\omega_b$ and $a_2$,   in the $B_s$ and two-pion distribution amplitudes. Therefore, we expect that the future measurement can   help us better  understanding the multi-body processes,  and S-wave two-pion resonance and $B_s$ distribution amplitudes.

\section*{Acknowledgments}
 This work is
supported in part by   National Natural Science Foundation of
China under Grant No.~11575110,  and  11735010, by Natural Science Foundation of Shanghai under Grant  No.~15DZ2272100,  by Key
Laboratory for Particle Physics, Astrophysics and Cosmology,
Ministry of Education.

\end{document}